\documentclass{article}

\usepackage{amsmath,amsthm,amssymb,amscd,dsfont}

\theoremstyle{plain}

\theoremstyle{definition}

\renewcommand{\S}{\mathcal{S}}

\newcommand{\SigObs}{\sigma_o}

\newcommand{\h}{\mathbf{h}}

\renewcommand{\t}{\mathbf{t}}
\newcommand{\nill}{\mathbf{0}}

\newcommand{\x}{{\mathbf{x}}}
\newcommand{\y}{{\mathbf{y}}}
\newcommand{\z}{\mathbf{z}}

\newcommand{\Id}{\mathds{1}}

\begin{document}

\title{Crystallographic Refinement using Non-Spherical Form Factors in \textit{olex2.refine}} 
\author{Laura Midgley\thanks{Department of Mathematical Sciences, Durham University, South Road, Durham, DH1 3LE, UK} \and  Luc J. Bourhis\thanks{Bruker France, 4 All\'ee Lorentz, Champs-sur-Marne, 77447 Marne-la-Vall\'ee cedex 2, France} \and Oleg Dolomanov\thanks{OlexSys, Chemistry Department, Durham University, South Road, Durham, DH1 3LE, UK} \and Norbert Peyerimhoff\footnotemark[1] \and Horst Puschmann\footnotemark[3]}
\maketitle

\begin{abstract}
We have implemented a procedure that allows the use of non-spherical atomic form factors in a standard crystallographic X-Ray refinement. We outline the procedure for their use, alongside a mathematical justification of their viability.
\end{abstract}

\section{Introduction}

We have implemented a procedure that allows the use of non-spherical atomic form factors in crystallographic refinement. It is agnostic to the method employed to compute those form factors, as the refinement engine \textit{olex2.refine} \cite{BDGHP14} will use tabulated atomic form factors if the file specified below is present. The refinement will proceed as expected within crystallographic refinement, including the ability to make use of constraints, restraints, disorder and other specific tools like twin refinement and solvent masking. This procedure is available from all versions of \textit{Olex2-1.3} \cite{CIT1}.

Crystallographic refinement typically treats atoms as isolated, stand-alone entities with a spherically symmetrical electron charge distribution. A non-spherical treatment arises naturally from the fact that the electron density distribution of an atom is influenced by its environment. Historically, Stewart derived non-spherical form factors for bonded hydrogen atoms and commented: ``By necessity, if not by choice, crystallographers have treated bonded atoms as point nuclei with a spherically symmetrical distribution of electron charge'' \cite{stewart1965coherent}. 

The non-spherical form factors of the individual atoms -- and their tabulation in the required format -- can be obtained by various means. A possible starting point could be a molecular quantum mechanical wave-function calculation, followed by its transformation into electron densities and then the partitioning into atomic contributions. Alternatively, the electron densities themselves can be approximated using data-based contributing fragments.

It is not the subject of this short note to discuss the details -- or indeed merits -- of any of these methods. We merely wish to announce the fact that refinement based on non-spherical form factors is now possible in \textit{olex2.refine} and to provide a rigorous mathematical justification for refinement using non-spherical form factors in this way. We would like to make it clear that those form factors are not refined by \textit{olex2.refine}. Only the usual parameters (positions, ADPs, occupancies, etc) are refined. These form factors can (and must) be externally recomputed after each series of refinement cycles, so that the form factor of each atom keeps matching the chemical environment as it changes during refinement.

\section{Tabulated Atomic Form Factors}

\textit{Olex2} expects a file called \textit{[name].tsc} (matching the \textit{.hkl} file name) containing the following information in order to use the external atomic form factors:

The header of the \textit{[name].tsc} file is free-format, as long as it contains the space-separated list of atom names in the `SCATTERERS:' line and finishes with `DATA:'. Any identifier must be followed by a colon. The identifiers may start with a space.


\begin{table}[h]
\begin{tabular}{ll}
TITLE: & \textit{optional title of the structure} \\
SYMM: & `expanded' \textit{or list of symmetries} \footnotemark[1]\\ 
AD: & TRUE or FALSE (anomalous dispersion) \\
SCATTERERS: & \textit{space-separated list of all atoms} \\
$ [ \text{ANYTHING} ] $ : & \textit{colon must be present} \\
DATA: & (denotes the end of the header) \\
\end{tabular}

\begin{tabular}{lllllll}
h & k & l & $A_1$ & $A_2$ & \dots & $A_N$ \\
$h_1$  & $k_1$  &  $l_1$ &  $f_1(h_1,k_1,l_1)$  &  $f_2(h_1,k_1,l_1)$  &  \dots  &   $f_N(h_1,k_1,l_1)$  \\
$h_2$  & $k_2$  &  $l_2$ &  $f_1(h_2,k_2,l_2)$  &  $f_2(h_2,k_2,l_2)$  &  \dots  &   $f_N(h_2,k_2,l_2)$  \\
\vdots  & \vdots  &  \vdots &  \vdots  &   \vdots &  \vdots   &  \vdots  \\
$h_m$  & $k_m$  &  $l_m$ &  $f_1(h_m,k_m,l_m)$  &  $f_2(h_m,k_m,l_m)$  &  \dots  &   $f_N(h_m,k_m,l_m)$ 
\end{tabular}
\end{table}

\footnotetext[1]{In either case, all symmetry equivalent Miller indices must be present in the DATA section. If a list of symmetry operators, expressed as rotation matrices (e.g.: 1 0 0 0 1 0 0 0 1;-1 0 0 0 1 0 0 0 -1) is provided, then the Miller indices must be ordered into corresponding blocks -- and each block must have symmetry equivalent indices in the same position in each block and generated by the corresponding matrices. This allows for more efficient calculations during the refinement. Otherwise, if SYMM has the value `expanded', the indices can be present in any order.}

As will be shown in the theory section, $f_j(h_i,k_i,l_i)$ is the form factor (Fourier transform of the electron density) of the atom $A_j$ calculated in a coordinate system obtained by translating the origin of the crystallographic axes to the centre of atom $A_j$, at $h_i, k_i, l_i$. Index $j \in (1,\dots ,N)$ should run over all unique atoms of the asymmetric unit, and $i \in (1,\dots ,m)$ should run over at least all reflections defined in the \textit{.hkl} file and any equivalents under symmetry.

The complex values $f_j(h_i,k_i,l_i)$ must be written as ``Re,Im'' - their real component followed by a comma followed by the imaginary component, with no spaces.

The format and information specified in the \textit{.tsc} files was motivated by the mathematical derivations presented in the next section.

\section{Theory}

We will explain the mathematics behind the use of non-spherical form factors and how \textit{olex2.refine} has been adapted to enable their use. We keep the notation close to the one used in \cite{BDGHP14}. We will first discuss the standard case and then briefly discuss the modifications needed for twinning. The non-spherical case remains very similar to the spherical case, with some critical differences which we will summarise at the end of this discussion.

\subsection{Monocrystals}

Our mathematical arguments focus on the necessary modifications concerning the treatment of the calculated structure factor, which we denote by $F(\h,\y(\x))$. Here $\h$ (a row vector) is a triplet of Miller indices, $\y$ comprises the crystallographic parameters (atomic positions and atomic displacement parameters (ADPs) and chemical occupancies) and the refinement is carried out with respect to potentially reduced parameters denoted by $\x$. The dependency of $\y$ on $\x$ is known analytically, and we emphasise this dependency by writing $\y$ as $\y(\x)$, which therefore embodies all constraints.

The structure factor is the sum of the individual contributions of all atoms, partitioned into atoms symmetry equivalent to representatives $A_j$, $j= 1,2,\dots,N$, in the asymmetric unit 
\begin{equation} \label{eq:struct-form-fact}
F(\h,\y(\x)) = \sum_{j=1}^N \underbrace{\sum_{(R|\t) \in \S} f_j^{(R|\t)}(\h,\y(\x))}_{\text{atoms equivalent by symmetry to $A_j$}},
\end{equation}
with $R$ the rotational part and $\t$ the translational part of the symmetry operation $(R|\t) \in \S$.

The representative atoms $A_j$ lie at fractional locations $\z_j$ (a column vector) with atomic vibration tensor $U_j$ (a $3\times 3$ symmetric matrix). This information is contained within the vector $\y(\x)$. For the chosen representative atom $A_j$ in the asymmetric unit, its individual contribution $f_j^{(\Id|\nill)}$ is given by \footnote[2]{`=:' means that the right hand side is defined by the left}
\begin{eqnarray} \label{eq:form-fact-id-decomp}
  f_j^{(\Id|\nill)}(\h,\y(\x)) &=& f_j(\h,\y(\x)) e^{-\h U_j \h^T}
                                   e^{i 2\pi \h \z_j} \nonumber \\
&=:& f_j(\h,\y(\x)) G_j(\h,\y_j(\x)).
\end{eqnarray}
Here $\y_j(\x)$ is the subset of parameters of the structure pertaining to the $j$-th atom (namely $U_j$ and $\z_j$).

The form factor of the atom $A_j$ is calculated in a coordinate system obtained by translating the origin of the crystallographic axes to the centre of atom $A_j$, with no change in orientation. The form factor $f_j(\h,\y(\x))$ is then the Fourier transform of the electron density $\rho_j$  of $A_j$. In contrast to the case of spherical form factors, this electron density can now depend on the whole structure whose information is given in $\y(\x)$, as non-spherical form factors take the dependence of the electron density of the surrounding atomic environment into account.

The relation between $f_j^{(R|\t)}$ for a general $(R|\t) \in \S$ and $f_j$ is then
\begin{eqnarray} \label{eq:form-fact-R-decomp}
  f_j^{(R|\t)}(\h,\y(\x)) &=& f_j(\h R,\y(\x)) e^{-\h R U_j R^T \h^T} e^{i 2\pi \h R \z_j} e^{i 2\pi \h \t} \nonumber \\
                   &=& f_j(\h R,\y(\x)) G_j(\h R,\y_j(\x)) e^{i 2\pi \h \t} \nonumber \\
  &=:& f_j(\h R,\y(\x)) G_j^{(R|\t)}(\h,\y_j(\x)).
\end{eqnarray}

Note that \textit{in the case of spherical form factors}, the functions $f_j$ do not depend on the structure information $\y(\x)$ and, additionally, we have $f_j(\h R) = f_j(\h)$  since $f_j(\h)$ does then not depend on the direction of $\h$ but only on $\h M^* \h^T$, where $M^*$ is the reciprocal metric matrix. This is generally not true in the case of non-spherical form factors.

The least square minimization in the refinement procedure requires derivatives of the structure factor with respect to the
components of $\x = (x_1,\dots,x_n)$. Since the structure factor is the above sum \eqref{eq:struct-form-fact}, we only need to consider the
derivatives of the individual terms $f_j^{(R|\t)}(\h,\y(\x))$.

Using the product rule, we have for the derivative (by dropping the arguments $\y(\x)$ for ease of reading)
\begin{equation} \label{eq:diffformfact}
\frac{\partial f_j^{(R|\t)}}{\partial x_k}(\h) =
\frac{\partial f_j}{\partial x_k}(\h R) G_j^{(R|\t)}(\h) +
f_j(\h R) \frac{\partial G_j^{(R|\t)}}{\partial x_k}(\h).
\end{equation}

As we differentiate with respect to $x_j$, partial derivatives $\frac{\partial y_i}{\partial x_j}$ will appear via the chain rule.


The differential $\frac{\partial f_j}{\partial x_k}(\h R)$ in the first term on the right hand side of \eqref{eq:diffformfact} is more difficult to treat due to the complexity of the involved derivations. We make the assumption that the effect of this term for the least square minimisation procedure is relatively minor and thus take it as zero. The errors introduced via this and other assumptions will escalate if the structure changes without frequent updating of the non-spherical form factors. The validity of this assumption is expected to assert itself through the experimental exploration of this refinement technique in the field.

Given these considerations, the contributions of all symmetry equivalent atoms in both the structure factor and its derivatives require only the table $f_j(\h)$ for the current structure information $\y(\x)$ for each single representative $A_j$ for each step of the refinement procedure.

The experimental input to the refinement is a list of $\h$, $F_o^2(\h)$ and $\SigObs(\h)$, where the last two items are respectively the measured intensities (scaled
and with absorption corrections) and its estimated standard uncertainty (the \textit{.hkl} file). Refinement is then a non-linear least squares fit of $|F(\h,\y(\x))|^2$ to $F_o^2(\h_j)$, for all $\h$. Precisely, the objective function to minimise is
\begin{equation} \label{eq:LSM0}
\x \mapsto \sum_{r=1}^m w(\h_r)\left(|F(\h_r,\y(\x))|^2 - F_o^2(\h_r)\right)^2,
\end{equation}
for $m$ measured reflections, where $w(\h_r)$ are suitable weights. For more detail, see Section 2 of~\cite{BDGHP14}. 

The set of Miller indices $\h$ required for the tabulated non-spherical form factors (the \textit{.tsc} file) should correspond to the set of measured Bragg reflections and their symmetry equivalents. 

\subsection{Twinning}

In the case of twinning, one needs to ensure that the set of Miller indices to be considered contains the measured Bragg peaks for all twin components and their symmetry equivalents.

The modification required to the least squares \eqref{eq:LSM0} is to replace each term $|F(\h_r,\y(\x))|^2$, $r=1,\dots,m$, by a combination over the contributing twin components indexed by $l$, namely
\begin{equation} \sum_{l=1}^{d_r}\alpha_{l}|F(\h_{r,l},\y(\x))|^2
\end{equation}
where $d_r\leq d$ is the number of contributing components (where $d$ is the total number of components), $\alpha_l$ the fraction of the crystal volume occupied by the $l$-th contributing twin domain to the reflection $h_r$, and $h_{r,l}$ is the corresponding Miller index of this twin component contributing to this reflection. Note by equations \eqref{eq:struct-form-fact} and \eqref{eq:form-fact-R-decomp}, the calculation of $F(\h_{r,l},\y(\x))$ requires the information of the non-spherical form factors for $\h_{r,l}$ and all its symmetry equivalents. 

For more information on the general twinning procedure, see Section 5 of~\cite{BDGHP14}.

Therefore, all types of twinning can be handled by providing as input, for each reflection $\h_r$, the corresponding Miller indices $\h_{r,l}$ of the contributing components $l \in \{1,\dots,d\}$, and the matching $F_o^2(\h_r)$ and $\SigObs(\h_r)$.

For the computation of structure factors using non-spherical form factors, only those Miller indices $\h_{r,l}$ and their symmetry equivalents are necessary. It is of course well known that in the case of (pseudo-)merohedral twinning, the Miller indices for a given $r$ and varying $l$ are related to each other by a twin law but this is only a special case of the general scheme we have just described: the calculation of form factors does not need to be aware of this detail.

\subsection{Summary}

Let us finally cover the relevant differences to be taken into account when working with non-spherical form factors:

\begin{itemize}
    \item[(i)]  Form factors associated to atoms (with the origin at their centre) are no longer real, but are usually complex-valued (as the electron densities are non-spherical).
    \item[(ii)] It is no longer the case that $f_j(\h R) = f_j(\h)$ for rotations $R$ associated to symmetry equivalent atoms in the unit cell.
    \item[(iii)] Due to the change in the shape of form factors under shifts, there appears an additional term in the derivative of $f_j^{(R|\t)}$  -- the first term on the right hand side of \eqref{eq:diffformfact}. We assume that this is negligible for sufficiently small shifts.
    \item[(iv)] The provided form factors must cover a greater variety of Miller indices than would be needed in the spherical case (due to (ii)). That is, form factors must be provided for all Miller indices $\h$ with recorded reflections and all symmetry equivalents $R\h$, for $(R,\t)$ appearing in $\S$ (see \eqref{eq:struct-form-fact}). This is also relevant for dealing with twin laws.
\end{itemize}

\bigskip

{\bf{Acknowledgement:}}
We gratefully acknowledge the support for Laura Midgley through the Intensive Industrial Innovation Programme run by Durham University and funded by the European Regional Development Fund (ERDF).

\end{document}